\documentstyle[12pt,epsfig,epsf]{article}
\newcommand{\bdm}{\begin{displaymath}}
\newcommand{\edm}{\end{displaymath}}
\newcommand{\dst}{\displaystyle}

\newcommand{\institute}[1]{\parbox{16cm}{%
\centering\normalsize \sl #1}}
\vspace{2cm}

\title{
\bf Static Potential in the SU(2)-Higgs Model and the Electroweak Phase
Transition\thanks{Talk given at the Central European Triangle Symposium on 
Particle Physics, June 19, 1999, Zagreb}} 
\author{
A.~Pir\'oth \\
\institute{Institute for Theoretical Physics, E\"otv\"os University,\\
H-1088 Budapest, Hungary }\\
}
\date{}
\begin{document}
\maketitle

\begin{abstract}
We present a one-loop calculation of the static potential in the SU(2)-Higgs 
model. The connection to the coupling constant definition used in 
lattice simulations is clarified. The consequences in comparing lattice 
simulations and perturbative results for finite temperature applications 
are explored.
\end{abstract}

\section{Introduction}

One of the most fundamental and intriguing problems in present-day particle physics
is the observed baryon asymmetry of the universe---that is the fact
that we do not
see hardly any antimatter around us, but we can see a large amount of matter, and a
vast amount of photons. 

One could be inclined to attribute this asymmetry to the initial conditions of
the big bang---but this being rather ad hoc, we shall focus on
dynamical generation of baryons. Another possibility could be arguing that the
entire universe has no net baryon number, just our region is more baryonic, while
other regions are more antibaryonic. However, from the boundary between two such
regions, characteristic photons would bring us information, and lacking this, we
can safely state that in our vicinity of at least $10^{13}$ solar masses there is
just matter. Since no known mechanism can separate matter and antimatter on such 
large scales, we shall assume that the universe is baryonic and this asymmetry was 
formed some time after the big bang. 

In 1967 Sakharov gave three necessary conditions for baryogenesis, which are
\begin{enumerate}
\item Baryon number violation---which is obvious
\item C and CP violation---otherwise the number of baryons and
antibaryons
generated would be equal
\item Departure from thermal equilibrium---since quantum numbers do not
change in
thermal equilibrium
\end{enumerate}   
Baryon number violating processes are known to be present in GUT's---and
GUT's are
still favoured candidates for baryogenesis. However, it came as a slight
surprise that even within the standard model there are such processes, the
nonperturbative sphaleron processes. The standard model being experimentally
verified with great precision, the investigation of the possibility of electroweak
baryogenesis is clearly of great importance.

The departure from thermal equilibrium can be realised during the electroweak phase
transition. This is customarily described by the effective potential.
\begin{figure}[h]
\begin{center}
\epsfig {file=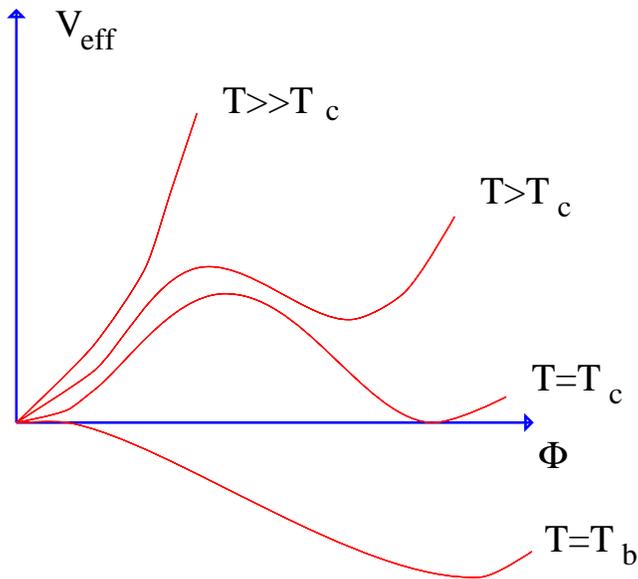, width=6cm, angle=270}
\vspace{2.5cm}
\caption{The effective potential as a function of the scalar field. The phase
transition point is defined by the degeneracy of the two minima}
\end{center}
\end{figure}

In order to have electroweak baryogenesis, the phase transition must be
strongly first order, or more quantitatively the relation
\bdm
\langle \Phi \rangle / T_C > 1
\edm
must hold, i.e.\ the vev of the Higgs field has to exceed the critical temperature.

How can we study this relation? 
The most straightforward method is resummed perturbation theory (cf.\ e.g.\ 
\cite{AE93,BFHW94,FH94}). In the low temperature Higgs phase
the perturbative approach is expected to work well, however, 
serious infrared problems are present in the high temperature symmetric phase.
Since the determination of thermodynamical quantities at the critical temperatures is
based on the properties of both phases, non-perturbative techniques are necessary for a
quantitative understanding of the phase transition.

One very successful possibility is to construct an effective 3-dimensional theory by 
using dimensional reduction, which is a perturbative step. The non-perturbative study
is carried out in this effective 3-dimensional model (see e.g.\ \cite{3d-sim}
and references therein). Analytical estimates are confirmed by numerical results 
and relative errors are believed to be at the percent level.

Another approach is to use 4-dimensional simulations. The complete lattice 
analysis of the standard model is not feasible due to the presence of chiral
fermions. however, the infrared problems are connected only with the bosonic
sector. These are the reasons why the problem is usually studied by simulating
the SU(2)-Higgs model on 4-dimensional lattices, and perturbative steps are used
to include the U(1) gauge group and the fermions.

Despite the fact that both perturbative and lattice approaches
are systematic and well-defined, it is not easy to compare
their predictions. The reason for this is that in lattice simulations
the gauge coupling constant is determined from the static potential, whereas
in perturbation theory the ${\overline {\rm {MS}}}$ scheme is used. 
Therefore, if one wishes to compare perturbative and lattice results,
a perturbative calculation of the static potential proves to be of great help. 

\section{Calculation of the one-loop static potential}

The concept of the static potential was introduced more than 20 years ago
by L.~Susskind \cite{Suss}.
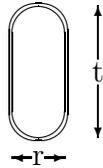
\begin{figure}[h]
\begin{center}
\begin{picture}(100, 60)(-40,-10)
\put(0,20){\oval (20, 50)}
\put(0,20){\oval (22, 52)}
\put(-2.5,-12.5) {\vector(-1,0){7.5}}
\put(2.5,-12.5) {\vector(1,0){7.5}}
\put(0,-12.5){\makebox (0,0){r}}
\put(22.5,15) {\vector(0,-1){20}}
\put(22.5,25) {\vector(0,1){20}}
\put(22.5,20){\makebox (0,0){t}}
\end{picture}
\caption{Heavy quark--antiquark loop}
\end{center}
\end{figure}

When a very heavy quark--antiquark pair is created from vacuum, 
separated and kept at distance $r$ for time $t$ and then let annihilate, the matrix 
element of the process can be given as
\bdm
\langle i | e^{-Ht} | f \rangle = e^{-V(r) t} \langle i|f \rangle,
\edm
and so
\begin{equation} \label{trans}
V(r) = - \lim_{t \to \infty}
\frac{1}{t} \log \langle W_{r,t} \rangle.
\end{equation}

The Feynman rules for these static sources are quite simple. Owing to their 
great mass, their coordinate space propagator is purely timelike, 

\epsfig{file=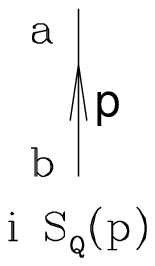, width=1.2cm}
\epsfig{file=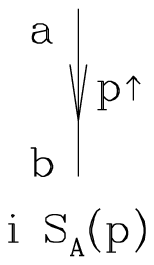, width=1.2cm}
\vspace{-2cm}

\bdm
i S_{Q/A}(x-y) = \delta ({\mathbf{x-y}})
\theta (\pm x_0 \mp y_0),
\edm
\bigskip

which upon Fourier transformation gives the momentum space propagator
\bdm 
\dst iS_{Q/A}^{ab}(p) \dst \frac{i}{v p +
i\epsilon},
\edm
where $v$ is the velocity of the source ([1,0,0,0] to a first approximation).

The heavy (anti)quark--gauge boson vertex is given by

\epsfig{file=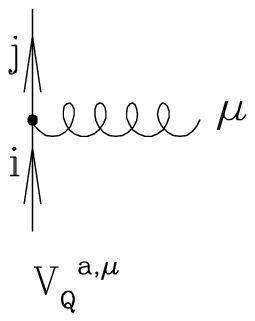, width=1.8cm}
\epsfig{file=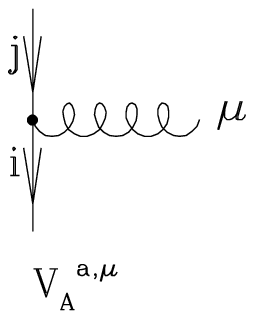, width=1.8cm}
\vspace{-2cm}
 
\bdm
\dst V_{Q/A}^{a,\mu} = \pm ig T^a_{i,j} \delta^{\mu 0}
\edm
\bigskip

By means of these rules, the one-loop static potential was calculated long ago in
quantum electrodynamics and quantum chromodynamics \cite{Suss,Fish,Appel}, and even
the full two-loop result was published recently \cite{Markus}.
The case of QED is rather simple: it was shown in \cite{Fish} that summing up all
orders in perturbation theory is equivalent to taking the exponential of the one
gauge boson exchange graph. However, this calculation is more than just a warm-up 
excercise, as the abelian parts of the graphs in more complicated theories (QCD,
SU(2)-Higgs model) are identical to those of QED. Therefore we can focus on graphs
with nonabelian contributions.

Our calculation was performed in the ${\overline {\rm {MS}}}$ scheme and  
the Feynman gauge but the result is gauge independent, as it should be for a
physical observable. The relevant graphs are shown in Fig.~3. 

Solid lines represent the heavy quark (antiquark) propagator, while 
wavy lines the vector boson propagator. External heavy quark 
(antiquark) propagators are not shown in the figure. The one-loop corrected 
vector boson propagator contains scalar and ghost contributions as well.
The result can be conveniently given in momentum space. One obtains
\cite{heg,Laine}

\begin{eqnarray}\label{mom_pot}
&\dst{V_{{\mathrm{1-loop}}}(k)=- \frac{3g^4}{32\pi^2}\frac{1}{k^2+M_W^2}} \nonumber \\
&\dst{\left\{\frac{k^2+M_W^2}{k}\frac{2}{\sqrt{k^2+4M_W^2}}
\log \frac{\sqrt{k^2+4M_W^2}-k}{\sqrt{k^2+4M_W^2}+k} + \right.} \nonumber \\
&\dst{\frac{1}{k^2+M_W^2} \left[\frac{1}{24R_{HW}^2}\left(86R_{HW}^2 k^2 
-9(6-3 R_{HW}^2 +R_{HW}^4 )M_W^2\right)\log\frac{\mu^2}{M_W^2} \right.} \nonumber \\
&\dst{+ \frac{1}{8} (13 k^2-20 M_W^2 ) F(k^2;M_W^2,M_W^2)}  \nonumber \\
&\dst{-\frac{1}{24}\left( (R_{HW}^2-1)^2 \frac{M_W^4}{k^2} +k^2 +2(R_{HW}^2-5)
M_W^2\right) F(k^2;M_W^2,M_H^2)} \nonumber \\
&\dst{+\frac{R_{HW}^2 \cdot \log R_{HW}}{12(R_{HW}^2-1)} \left( k^2+(9R_{HW}^2-17)M_W^2\right)} \nonumber \\
&\dst{\left. \left. +\frac{1}{72R_{HW}^2}\left( R_{HW}^2 k^2+3 (-18+R_{HW}^2-11 R_{HW}^4)M_W^2\right) \right] 
\right\} },
\end{eqnarray}
where $k^2$ denotes the square of the three-momentum $\vec{k}$, $M_H$ the 
Higgs mass and  $R_{HW}=M_H /M_W $. The function $F$ is defined as

\begin{eqnarray}\label{fv}
F(k^2;m_1^2,m_2^2)=1+\frac{m_1^2 +m_2^2 }{m_1^2 -m_2^2 } \log \frac{m_1}{m_2} +
\frac{m_1^2 -m_2^2 }{k^2} \log \frac{m_1}{m_2} \nonumber \\
+\frac{1}{k^2} \sqrt{ (m_1 +m_2 )^2 +k^2 )((m_1 -m_2 )^2 +k^2 )} \log 
\dst{\frac{1-\sqrt{\frac{(m_1 -m_2 )^2 +k^2}{(m_1 +m_2 )^2 +k^2 }}}
{1+\sqrt{\frac{(m_1 -m_2 )^2 +k^2}{(m_1 +m_2 )^2 +k^2 }}}}.
\end{eqnarray}
Although the dependence on the renormaization scale can be removed by 
introducing the one-loop W mass \cite{Laine}, we do not follow this line.

Eq.\ (\ref{mom_pot}) has to be Fourier transformed into coordinate space. 
We applied brute force methods performing numerical integration. As 
a check, we compared our results with various pieces of the partly analytic 
calculation in \cite{Laine} for the derivative of the potential 
(with respect to distance). The agreement is excellent.

\begin{figure}
\begin{center}
\epsfig{file=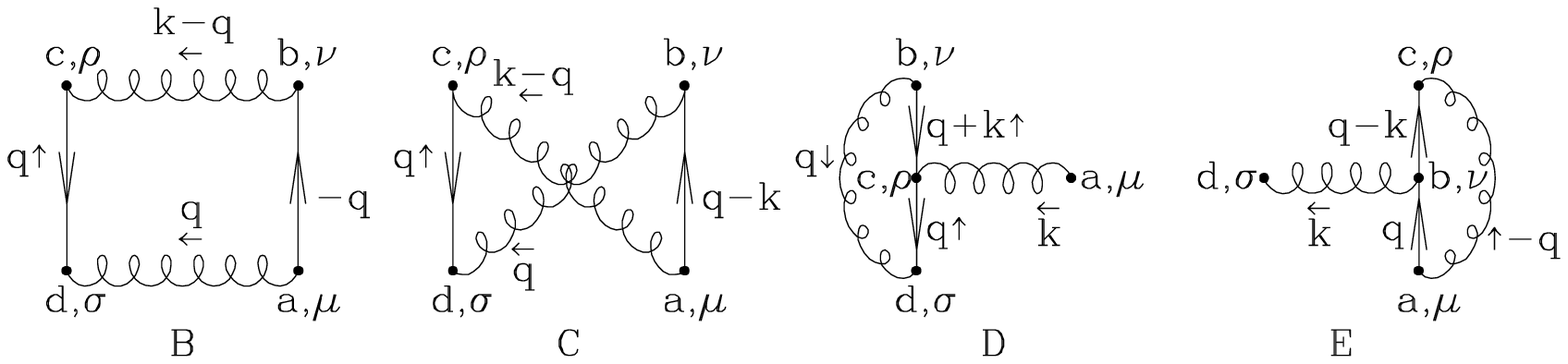,width=12.0cm}
\epsfig{file=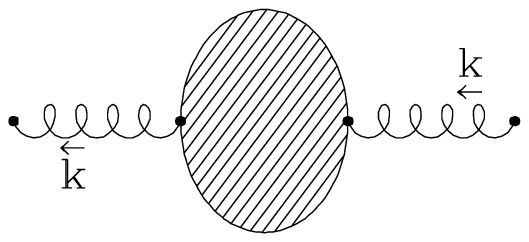,width=3.8cm}
\caption{\label{fig1}
{Graphs giving nonvanishing contributions to the static potential 
}}
\end{center}\end{figure}
Our result is presented in Figs.\ 2 and 3, where the various parts of the 
one-loop correction to the potential are  plotted. We define

\begin{eqnarray}\label{pot}
\frac{V(r)}{M_W}=-\frac{3g^2}{16\pi} \frac{\exp(-M_W^0 r)}{M_W r} + 
\frac{g^4}{16\pi^2} 
\left(A+B\log(\mu^2/M_W^2)\right),
\end{eqnarray}
where $M_W^0 =M_W-\delta M_W$, with $\delta M_W$ the one-loop mass correction. 
Since $\delta M_W$ is scale dependent, so is $M_W^0$.
A and B are functions of the distance $r$ and $R_{HW}=M_H/M_W$. 
We choose $M_W=80$GeV.
Fig.~4 shows the  dependence of $A$ and $B$ on the dimensionless distance 
$rM_W$ for $R_{HW}=0.8314$ (corresponding to the end point of the first 
order finite  temperature phase transition \cite{Fod8}), while Fig.~5 
shows the $R_{HW}$ dependence for $r=M_W^{-1}$.

\begin{figure}
\begin{center}
\epsfig{file=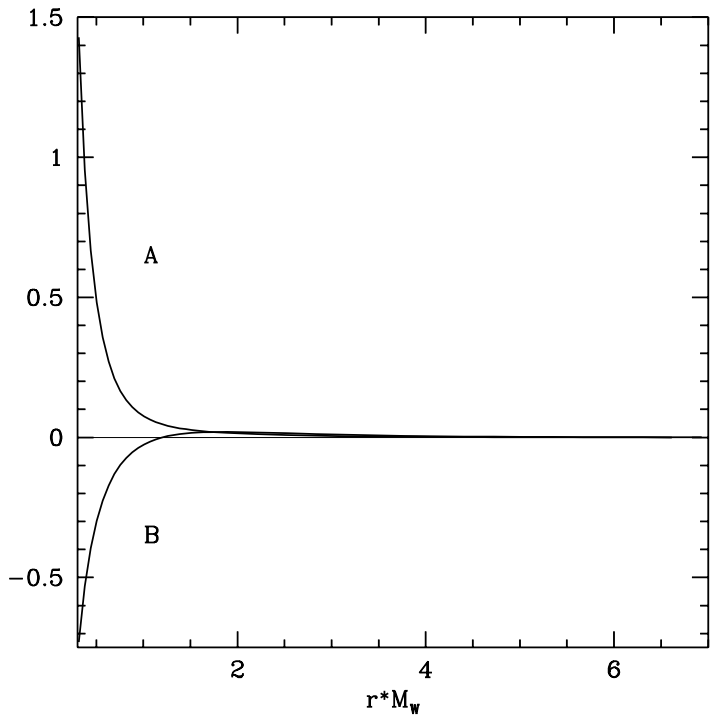,width=10.0cm}
\caption{\label{pot_fig}
{The coefficients of $g^4/(16 \pi^2)$---curve A---and of $g^4/(16\pi^2) 
\log(\mu^2/M_W^2)$---curve B---of the static potencial Eq.\ (\ref{pot}) as a function of 
distance times W mass. $R_{HW}$=0.8314.
}}
\end{center}\end{figure}

\begin{figure}
\begin{center}
\epsfig{file=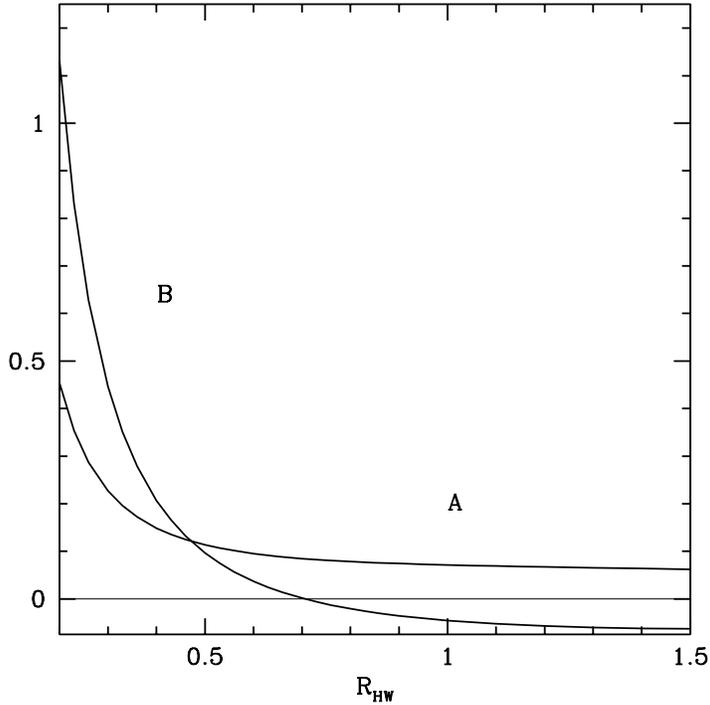,width=10.cm}
\caption{\label{pot_R_fig}
{The coefficients of $g^4/(16\pi^2)$---curve A---and of $g^4/(16\pi^2) 
\log(\mu^2/M_W^2)$---curve B---of the static potencial Eq.\ (\ref{pot}) 
as a function of $R_{HW}$. 
The distance is $M_W^{-1}$.
}}
\end{center}\end{figure}
\section{Relation of the continuum version of the lattice coupling constant
definition to the ${\mathbf{\overline {\mathbf {MS}}}}$ coupling constant }

Since we wish to compare results of lattice simulations and continuum perturbation theory 
calculations, it is an essential point to define the SU(2) gauge coupling in the 
same way in both cases. However, in continuum perturbation theory the 
$\overline {\rm {MS}}$ running coupling constant at a given renormalization scale 
is more natural (as used in Eqs.\ (\ref{mom_pot},\ref{pot}), too), while in 
lattice simulations other definitions are applied. 
Therefore we have to establish the relation between the coupling constants.

The lattice definition of the coupling constant (inspired by \cite{Som}) is given in \cite{Fod1}. 

First rectangular Wilson loops of size $(r,t)$ are measured. 
Extrapolating to large $t$ and dividing the logarithm by $-t$ one gets
the static potential in the $t \rightarrow \infty$ limit as a function of $r$ [see 
Eq.\ (\ref{trans})].
The nonperturbative lattice static potential is fitted by a finite lattice 
version of the Yukawa potential with four parameters (for details cf.\ 
\cite{Fod1}). One of these parameters is the mass in the exponential of the 
Yukawa potential, which is usually called the screening mass. The gauge 
coupling  at distance $r$ is defined as the ratio of the discrete 
$r$ derivative of the lattice simulated nonperturbative potential 
and the discrete derivative of the tree-level 
lattice Yukawa potential normalized by the square of the tree-level coupling 
and with the mass parameter $M_{\mathrm{lattice}}$ identified with the screening mass.
In practice $g^2_{\mathrm{lattice}} (M_{\mathrm{lattice}}^{-1} )$ is determined
and is called the local renormalized gauge coupling constant on the lattice.
The lattice results at various Higgs masses are collected in Table 1.
Data are from \cite{Fod8, Fod1, 4d-asym, Fod3}.

The gauge coupling constant can be defined in the same spirit in the case of
continuum perturbation theory, too:
\begin{equation}\label{g2_def}
g_{R}^2 (r)=\frac{1}{C_F}\frac{\dst\frac{d}{dr}\left[-V(r)\right]}
{\dst\frac{d}{dr} \int \frac{d^3 k}{(2\pi)^3}\frac{\exp (i{\vec k}\cdot {\vec r})}
{k^2 + M_{\rm screen}^2 }},
\end{equation}
i.e.\ by taking the ratio of the derivatives with respect to 
$r$ of the one-loop potential and the tree-level potential normalized by the 
square of the tree-level coupling. In Eq.\ (\ref{g2_def}) $V(r)$ is given by Eq.\
(\ref{pot}), $C_F=3/4$, and $M_{\rm screen}$ is obtained  from the fit.
 
Since $M_{\rm screen}-M_W^0=O(g^2)$, for distances satisfying 
$M_{\rm screen} -1/r =O(g^2) $ we 
can put Eq.\ (\ref{g2_def}) into the form
\begin{eqnarray}\label{g2_res}
g_{R}^2 (r)= g_{\overline {\rm {MS}}}^2 (\mu)\left(1+\frac{1}{2}
\left(1-\frac{M_W^0}{M_{\rm screen}}\right)\right)+\frac{g_{\overline {\rm {MS}}}^4 (\mu)}
{16\pi^2} \left(C+D \log \frac{\mu^2}{M_W^2} \right).
\end{eqnarray}
$C$ and $D$ are functions of $R_{HW}$ and $M_{\rm screen}$, their 
values are tabulated in Table 2 for $M_{\rm screen}=M_W=80$GeV.

\begin{table}[htb]
\begin{center}
\begin{tabular}{|c|c|c|c|c|}
\hline
 $R_{HW}$ & .2049 & .4220 & .595  & .8314 \\
 \hline
 $T_C$  (GeV)  & 38.3 & 72.6 & 100.0 & 128.4\\
 \hline
 $M_{\mathrm{lattice}}$ (GeV)& 84.3(12) & 78.6(2) & 80.0(4) & 76.7(24) \\
 \hline
 $g^2_{\mathrm{lattice}} (M^{-1} )$ & .5630(60) & .5788(16) & .5782(25) & .569(4) \\
 \hline
 $M_{\rm screen}$ (GeV) &74.97 & 80.44 & 80.70 & 81.77 \\
 \hline
 $g^2_{\overline {\rm {MS}}} (T_C )$ &0.540 & 0.592 & 0.585 & 0.570 \\
 \hline
 \end{tabular}
 \caption{\label{couplings}
 Various quantities calculated for values of $R_{HW}$ used in lattice 
 simulations.
 }
 \end{center}
 \end{table}

In this procedure we have to choose the  gauge coupling in the one-loop potential
so that $g_R^2 (M_{\rm screen}^{-1} )$ reproduces the lattice result (third row of 
Table 1) for the appropriate value of the Higgs mass. 
For the applications of the following section  (thermodynamical quantities at and
around the critical temperature $T_C$ of the first order electroweak phase transition)
the scale of the one-loop potential is chosen to be $T_C\approx 2M_H$, where 
$M_H$ is the Higgs boson mass at zero temperature.  
 Thus the  
gauge coupling appearing in the one-loop potential is actually the 
${\overline {\rm {MS}}}$ gauge coupling 
at scale $T_C$. The ${\overline {\rm {MS}}}$ gauge coupling values obtained from this 
procedure are given in the sixth row of Table 1.

Other definitions of the perturbative gauge coupling are also possible \cite{Laine}; 
however, our definition seems to provide the most systematic way of comparing
perturbative and nonperturbative results.

\begin{table}[htb]
\begin{center}
\begin{tabular}{|c|c|c|}
\hline
 $R_{HW}$ & $C$ & $D$ \\
 \hline
 0.2 &-41.54 & -22.19\\
 \hline
 0.3 &-8.26 & -6.58\\
 \hline
 0.4 &-6.47 & -1.12\\
 \hline
 0.5 & -5.66& 1.39\\
 \hline
 0.6 & -5.23& 2.74\\
 \hline
 0.7 &-4.98 & 3.55\\
 \hline
 0.8 & -4.83& 4.06\\
 \hline
 0.9 &-4.72 & 4.39\\
 \hline
 1.0 & -4.65& 4.62\\
 \hline
 1.1 & -4.59& 4.78\\
 \hline
 1.2 & -4.54& 4.89\\
 \hline
 1.3 & -4.50& 4.98\\
 \hline
 1.4 & -4.45& 4.98\\
 \hline
 1.5 &-4.40 & 5.01\\
 \hline
 \end{tabular}
 \caption{\label{coupling_res}
 The coefficients $C$ and $D$ defined in
 Eq.\ (\ref{g2_res}) as a function of $R_{HW}$.
 }
 \end{center}
 \end{table}

\section{Comparison of perturbative and lattice results for
physical observables}

In this section we compare lattice results and perturbative predictions for the
finite temperature electroweak phase transition. 

Lattice Monte Carlo simulations provide a 
well-defined and systematic approach to study the features of the finite
temperature electroweak phase transition.
During the last years large scale numerical simulations have been carried out
in four dimensions in order to clarify non-perturbative details
\cite{4d-asym},\cite{Fod8},\cite{Fod1},\cite{Fod3}.    
Thermodynamical quantities (e.g.\ critical
temperature, jump of the order parameter, interface tension, latent heat)
have been determined and extrapolation to the continuum limit has been performed in
several cases. Nevertheless, it has proven difficult
to compare perturbative and lattice results, because the
perturbative approach used the ${\overline {\rm {MS}}}$ scheme for the gauge coupling,
whereas the lattice determination of the gauge coupling has been based on
the static potential. This difference between the definitions can be removed on the basis of the
previous section. 

In this paper we use the published perturbative two-loop result for the
finite temperature effective potential of the SU(2)-Higgs model
\cite{FH94}. Note that the numerical evaluation of
the one-loop temperature integrals gives a result which agrees with the
approximation based on high temperature expansion within a few percent.
The reason for this is that the perturbative expansion
up to order $g^4,\lambda^2$ corresponds to a high temperature
expansion, which is quite precise for the Higgs boson masses we studied.
It is known that the perturbative loop expansion becomes unreliable
for Higgs masses above approximately 50 GeV (e.g.\ resummed perturbation 
theory fails to predict the end-point of the electroweak phase transition, thus it
gives a first order phase transition for arbitrarily large
Higgs boson masses). In the physically relevant range of the parameter space the
electroweak phase transition can only be understood by means of
non-perturbative methods. Therefore it is particularly
instructive to see quantitatively how perturbative and lattice results 
agree for small Higgs boson masses and how they differ for larger ones.

In lattice simulations masses are extracted from correlation functions, and it
is possible to use the zero temperature effective potential in order to
include the most important mass renormalization effects. 
The Higgs boson
mass obtained from the asymptotics of the correlation function corresponds 
to the physical mass determined by the pole of the propagators, i.e.\ the 
solution of $p^2-M^2=\Pi (p^2)$, where $\Pi (p^2)$ is the self-energy. The
effective potential approach
suggested by Arnold and Espinosa \cite{AE93} approximates $\Pi(p^2)$ by
$\Pi (0)$ in the above dispersion relation. It has been argued that the
difference between the two expressions is of order $g^5 v^2$  ($v$ is the
zero-temperature vacuum expectation value), which
does not affect our discussion. In this scheme the correction to the
$\overline{\rm {MS}}$ potential reads
\begin{equation}
\delta V={\varphi^2 \over 2} \left( \delta m^2+ {1 \over 2\beta^2}
\delta\lambda\right) + {\delta\lambda \over 4} \varphi^4,
\end{equation}
where
\begin{equation}
\delta m^2 = {9g^4v^2 \over 256 \pi^2},\ \ \ \ \ \
\delta\lambda=-{9g^4\over 256\pi^2}\left(\log\frac{M_W^2}{\mu}+{2\over
3} \right).
\end{equation}
Here ${\mu}$ is the renormalization scale and $M_W$ is the W-boson
mass at $T=0$. The above
notation corresponds to a tree-level potential of the form
$m^2 \varphi^2/2+\lambda \varphi^4/4$. Note that this treatment is
analogous to previous comparisons of perturbative and lattice results
\cite{BFH95}.

In \cite{Fod8, Fod1, 4d-asym, Fod3} several observables 
were determined, including renormalized masses at zero temperature ($M_H$,
$M_W$), critical temperatures ($T_C$), jumps of the order parameter 
($\varphi_+$), latent heats ($Q$) and surface tensions ($\sigma$) for
different Higgs boson masses. As usual, the dimensionful quantities were
normalized by the proper power of the critical temperature. Simulations
were performed on $L_t=2,3,4,5$  lattices ($L_t$ is the temporal
extension of the finite-temperature lattice) and whenever
it was possible a systematic
continuum limit extrapolation was carried out assuming standard $1/a^2$
corrections for the bosonic theory.

\begin{table}  \begin{center}
\label{comp}
\begin{tabular}{|c|}
\hline
 $M_H$\\ \hline $g_R^2$  \\
 \hline
  \begin{tabular}{c|c}
  $T_C/M_H$ \hspace{-5.5pt} &
   \begin{tabular}{c}
   pert \\
   \hline
   nonpert
  \end{tabular}
 \end{tabular} \\
 \hline
  \begin{tabular}{c|c}
 $\varphi_+/T_C$ \hspace{-1.5pt} &
  \begin{tabular}{c}
   pert \\
   \hline
   nonpert
  \end{tabular}
 \end{tabular} \\
 \hline
  \begin{tabular}{c|c}
  $Q/T_C^4$ \hspace{1.5pt} &
   \begin{tabular}{c}
   pert \\
   \hline
   nonpert
   \end{tabular}
  \end{tabular} \\
\hline
\begin{tabular}{c|c}
$\sigma/T_C^3$ \hspace{3.6pt} &
\begin{tabular}{c}
pert \\
\hline
nonpert
\end{tabular}
\end{tabular}\\
\hline
\end{tabular}
\begin{tabular}{|c|c|c|c|}
\hline
16.4(7)          & 33.7(10)         & 47.6(16)         & 66.5(14) \\
\hline
0.561(6)         & 0.585(9)         & 0.585(7)         & 0.582(7) \\
\hline
2.72(3)          & 2.28(1)          & 2.15(2)          & 1.99(2)  \\
\hline
2.34(5)          & 2.15(4)          & 2.10(5)          & 1.93(7)  \\
\hline
4.30(23)         & 1.58(7)          & 0.97(4)          & 0.65(2)  \\
\hline
4.53(26)         & 1.65(14)         & 1.00(6)          & 0 \\
\hline
0.97(7)          & 0.22(2)          & 0.092(6)         & 0.045(2) \\
\hline
1.57(37)         & 0.24(3)$^*$      & 0.12(2)          & 0 \\
\hline
0.70(10)         & 0.067(6)         & 0.022(2)         & 0.0096(5) \\
\hline
0.77(11)         & 0.053(5)$^*$    & 0.008(2)$^*$     & 0 \\
\hline
\end{tabular}
\end{center} 
\caption[]{ Comparison of the perturbative and the lattice results.}
\end{table}

The statistical errors of these observables are normally determined by
comparing statistically independent samples. 
Jackknife and bootstrap techniques were used \cite{14} and correlated 
fits were performed \cite{15} to obtain reliable estimates of the statistical 
uncertainties. 
A correct comparison has to include errors on the parameters used in the perturbative
calculation. These uncertainties are connected with the fact that neither the
Higgs boson mass nor the gauge coupling constant
can be   determined exactly in lattice simulations. Including  these errors, 
the perturbative prediction for an observable is rather an interval than one 
definite value. 

To obtain a better measure of the correspondence between perturbative and 
nonperturbative results, and to incorporate their errors, one introduces ``pulls''
defined by the expression
\begin{equation}
\mathrm{pull} = \frac{{\mathrm{perturbative\ mean}} - {\mathrm{nonperturbative\ mean}}}
{{\mathrm{perturbative\ error}} + {\mathrm{nonperturbative\ error}}}.
\end{equation}
The four different pulls at different Higgs boson masses are tabulated in Table 4 and
plotted in Fig.~6. 
For the sake of convenience, we used the shorthand $P_T=\mathrm{pull\ of\ } T_C/M_H$, 
$P_\phi=\mathrm{pull\ of\ } 
\varphi_+/T_C$, $P_Q=\mathrm{pull\ of\ } Q/T_C^4$, and $P_\sigma=\mathrm{pull\ of\ }
\sigma/T_C^3$.
\begin{table}[htb]
\begin{center}
\begin{tabular}{|c|c|c|c|c|}
\hline
$m_H$ (GeV) & 16.4(7) & 33.7(10) & 47.6(16) & 66.5(14) \\
\hline
$P_T$       & 4.75    & 2.60     & 0.71     & 0.67 \\
\hline
$P_\varphi$ & 0.47    & -0.33    & -0.3     & 32.5 \\
\hline
$P_Q$	    & -1.36   & -0.4    & -1.08    & 22.5 \\
\hline
$P_\sigma$  & -0.33   & 1.27	 & 3.5	    & 19.2 \\
\hline
\end{tabular}
\caption{Values of the four different pulls for various Higgs boson masses}
\end{center}
\end{table}
\begin{figure}
\begin{center}
\hspace{-1cm}
\epsfig{file=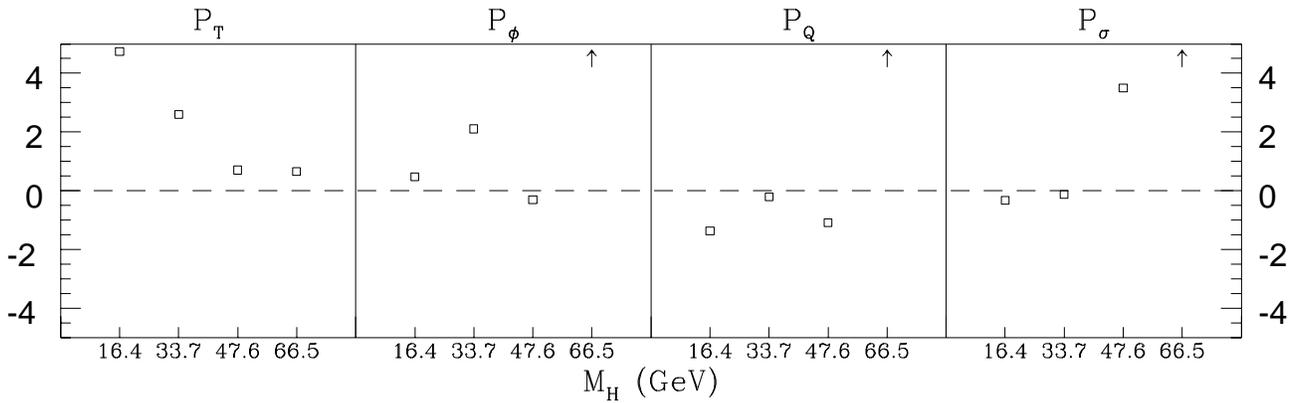,width=18cm}
\caption{\label{fig4}
``Pulls'' plotted against the Higgs mass. 
Arrows indicate values outside the
interval $[-5,5]$.}
\end{center}
\end{figure}

The quantity which has the smallest pull even for large Higgs boson masses 
is $T_C /M_H $. A quadratic fit was performed to this quantity as a function 
of $R_{HW}$. The result is
\begin{equation}
\frac{T_C}{M_H}=2.494-0.842 R_{HW} + 0.223 R_{HW}^2.
\end{equation}

For large Higgs masses he unreliability of perturbative predictions (in
particular concerning the quantity $P_\phi$, which is of central importance
from the viewpoint of baryogenesis) is striking.

\section{Conclusions }

Searching for the origin of the observed baryon asymmetry of the universe one
should focus on the electroweak phase transition. By calculating the 
SU(2)-Higgs static potential perturbatively, one can establish a better
connection between perturbative and nonperturbative studies of the
electroweak phase transition. 

From this relation it can be seen that a purely perturbative study is not
satisfactory. One also comes to the conclusion that electroweak baryogenesis
is ruled out in the standard model---or to put it rather positively:
there is 
physics beyond the standard model. 

Therefore, extensions of the standard model have to be studied---preferrably
on the lattice, either in dimensionally reduced theories or in 4D. A very promising
candidate is the MSSM, although the large number of free parameters makes this 
study a formidable task. However, within the MSSM the baryon asymmetry
can be accounted for only if the lightest Higgs mass is $M_H < 115
\mathrm{GeV}$ (for a recent review, see \cite{quiros}). Therefore experiments 
will either rule out this scenario in the close future, or, measuring several 
parameters, will facilitate numerical simulations. 

\section*{Acknowledgements}

I would like to thank F.~Csikor, Z.~Fodor, P.~Heged\"{u}s and 
$\Gamma$.~Koutsoumbas for useful discussions. \medskip

This work was partially supported by
Hungarian Science Foundation Grants under Contract  
No.\ OTKA-T22929-29803-M28413/FKFP-0128/1997.

\vfill

\end{document}